\newcommand{\q}{\label}
\newcommand{\pl}{\partial}
\newcommand{\be}{\begin{equation}}\newcommand{\ee}{\end{equation}}
\newcommand{\bea}{\begin{eqnarray}}\newcommand{\eea}{\end{eqnarray}}
\newcommand{\nn}{\nonumber}\newcommand{\p}[1]{(\ref{#1})}
\newcommand{\un}{\underline}
\documentstyle[12pt]{article}
\begin{document}

\thispagestyle{empty}
\begin{flushright}
BONN-HE-92-19\\ JHU-TIPAC-920018\\ ENSLAPP-L-392-92
\end{flushright}

\begin{center} {\bf
\Large{\bf  A twistor formulation of the heterotic D=10 superstring
with manifest (8,0) worldsheet supersymmetry}}\end{center} \vskip
1.0truecm

\centerline{{\bf F. Delduc${}^{(a)}$,
A. Galperin${}^{(b)*\dag}$, P. Howe${}^{(c)}$ and E.
Sokatchev${}^{(d)**}$}}
\vskip5mm

\centerline{${}^{(a)}$ \it Lab. de Phys. Th\'eor. ENSLAPP, ENS Lyon}
\centerline{ 46 All\'ee d'Italie, 69364 Lyon, France}
\vskip5mm
\centerline{${}^{(b)}$\it Department of Physics and Astronomy }
\centerline{\it  Johns Hopkins University, Baltimore, MD 21218, USA}

\vskip5mm
\centerline{${}^{(c)}$ \it Department of Mathematics, King's College,
London, UK}
\vskip5mm
\centerline{${}^{(d)}$\it Physikalisches Institut, Universit\"at Bonn}
\centerline{\it Nussallee 12, D-5300 Bonn 1, Germany}
\vskip 1.0truecm  \nopagebreak

\begin{abstract}
We propose a new formulation of the heterotic $D=10$ Green-Schwarz
superstring whose worldsheet is a superspace with two even and eight odd
coordinates. The action is manifestly invariant under both target-space
supersymmetry and a worldsheet reparametrisation supergroup.  It contains
only a finite set of auxiliary fields.  The key ingredient are the
commuting spinor (twistor) variables, which naturally arise as worldsheet
superpartners of the target space Grassmann coordinates. These spinors
parametrise the sphere $S^8$ regarded as a coset space of the $D=10$
Lorentz group.  The sphere is associated with the lightlike vector of one
of the string Virasoro constraints. The origin of the on-shell $D=10$
fermionic kappa symmetry of the standard Green-Schwarz formulation is
explained. An essential and unusual feature is the appearance of the
string tension only on shell as an integration constant.
\end{abstract}
\vskip6mm \nopagebreak \begin{flushleft} \rule{2 in}{0.03cm} \\
{\footnotesize \ ${}^{*}$ On leave from the Laboratory of Theoretical
Physics, Joint Institute for Nuclear Research, Dubna, Russia}\\
{\footnotesize \ ${}^{\dag}$ This work has been supported by the U.S.
National Science Foundation, grant PHY-90-96198}
\\ {\footnotesize \ ${}^{**}$  On leave from the
Institute for Nuclear Research and Nuclear Energy, Sofia, Bulgaria}
\vskip8mm
July 1992 \end{flushleft}

\newpage\setcounter{page}1

\section{Introduction}

The superstring is a very geometrical theory which has two apparently
unrelated presentations, the worldsheet supersymmetric
Neveu-Schwarz-Ramond version and the spacetime supersymmetric
Green-Schwarz version, which nevertheless turn out to be equivalent (see
\cite{GSW} for a review). An
interesting problem, and the subject of this paper, is to construct a
version of the theory which has both worldsheet and space-time
supersymmetry built in.  To some extent the Green-Schwarz superstring does
have this feature, since the GS action is invariant under a local
fermionic symmetry, kappa-symmetry \cite{S}, which plays a crucial r\^ole.
This feature of string theory occurs also in superparticles of
Brink-Schwarz type \cite{BS}, but until recently, there was no geometrical
understanding of kappa-symmetry. In reference \cite{STV} Sorokin, Tkach,
Volkov and Zheltukhin were able to reformulate the superparticle (in $D=3$
and $4$ dimensions) in a way in which this symmetry was interpreted as
worldline supersymmetry ($N=1$ and 2, respectively).  Their reformulation
involved the introduction of twistor-like variables which arise
essentially from solving the masslessness constraint $p^2=0$.
Subsequently this formalism was extended to cover the $D=6$ case ($N=4$
worldline supersymmetry) and also general target superspaces \cite{DS}.
In \cite{HT} the topological nature of the STVZ action was emphasised: it
is invariant under a restricted, but non-trivial group of worldline
diffeomorphisms even though it involves no worldline ``supergravity''
fields. A group-theoretical analysis of the twistor-like variables for the
cases $D=3,4,6$ and $10$ was given in
\cite{GHS}, \cite{HW}. The
idea of ref. \cite{STV} has been generalised to cover superstrings in
$D=3,4,6$ dimensions in refs. \cite{B1}, \cite{IK}, \cite{DIS}. These
string actions can be characterised as being ``semi-topological'' in that
they resemble the superparticle actions with respect to left
diffeomorphisms, but worldsheet supergravity fields are necessary in the
right sector.  Note that in the cases $D=4$ and $D=6$ the notions of
complex and quaternionic structures and the associated Grassmann
analyticity (chirality in $D=4$ and harmonic analyticity in $D=6$) were
heavily used both for the superparticle \cite{STV}, \cite{DS} and for the
superstring
\cite{IK}, \cite{DIS}. Here one should also mention  formulations of
the $D=10$ superstring with only $N=2$ manifest worldsheet supersymmetry
\cite{TON1}, \cite{B2}.

In this paper we shall present an action for the heterotic superstring in
10 dimensions and show its classical equivalence (after eliminating a
certain number of auxiliary fields and fixing some gauges) to the usual
Green-Schwarz action.  The new action is an integral over an $N=(8,0)$
worldsheet superspace and at the same time has a $D=10$ target superspace.
Its manifest local worldsheet supersymmetry replaces kappa-symmetry of the
usual superstring.  We emphasise that the case $D=10$ differs from
$D=3,4,6$ even for the superparticle, largely because in $D=10$ there is
no obvious replacement for the complex and quaternionic structures present
in $D=4$ and $6$, respectively. A way round this difficulty was recently
found in ref.
\cite{GS}. There an $N=8$ worldline supersymmetric action for the
$D=10$ superparticle was given which leads to the $N=8$ worldline
supersymmetric equations proposed in
\cite{GHS}. In ref. \cite{GS} the way to generalise the approach to the
superstring was indicated.  In this paper we complete the development of
the ideas proposed in \cite{GHS} and \cite{GS}.  \footnote{Recently
M.Tonin sent us his preprint
\cite{TON}, where a formulation of the $D=10$ superstring is proposed. In
some aspects it resembles ours, and later on in the text we shall comment
on that.} We shall focus on the $D=10$ case, but, as in the case of the
superparticle, the formalism is also applicable to $D=3,4,6$.

The key ingredients in the construction are the twistor-like variables
which parametrise the celestial sphere $S^8$ and the appearance of the
string tension as a cohomological term (an integration constant) in a
Lagrange multiplier superfield. Both of these are also features of the
superparticle action given in
\cite{GS}.  \footnote{The idea to
obtain the string tension as an integration constant was first proposed in
\cite{T} in the context of Green-Schwarz type actions without worldsheet
supersymmetry.  Later on it was generalised to $p$-branes \cite{BT}. In
this paper we shall show that worldsheet supersymmetry does in fact {\it
require} such an unusual way to introduce the string tension. } The
twistor-like variables arise as natural superpartners of the Grassmann
target space coordinates $\theta$ with respect to worldsheet
supersymmetry. Geometrically, the construction can be viewed in the
following way: the string can be thought of as a $(2\vert 8)$ (2 even, 8
odd-dimensional) subsupermanifold, ${\cal M}$, of the $(10\vert 16)$ super
target space, $\un{\cal M}$, embedded in such a way that the odd part of
the tangent space to ${\cal M}$ lies entirely within the odd part of the
tangent space to $\un{\cal M}$ at any point of ${\cal M}$. Thus, at each
point in ${\cal M}$ an eight-dimensional subspace of the odd tangent space
to $\un{\cal M}$ is selected, and the set of such subspaces is
parametrised by the coset space $Spin(1,9)/H$, where $H$ is the Borel
subgroup which takes a reference subspace into itself. This coset space is
the eight-sphere \cite{GHS}. The string tension arises from the term in
the action involving the supergravity two-form $B$.  This part of the
action is ``almost'' topological: variation with respect to the Lagrange
multiplier field $P$ sets the pull-back of $B$ onto the worldsheet to be
almost pure gauge (in fact, it is pure gauge in the left directions, which
is a reflection of lightlike integrability). The Lagrange multiplier
itself possesses a large abelian gauge invariance which, together with the
equations of motion implies that only a single constant component survives
in it. This constant is identified with the string tension.

We emphasise that the treatment of the superstring given in this paper is
purely classical. Quantisation of twistor-like theories is a non-trivial
subject, which deserves a separate study and will not be addressed here.

The paper is organised as follows. In Section 2 we explain how the bosonic
massless particle in $D=10$ can be formulated in terms of the new twistor
variables. Unlike the original approach of \cite{STV}, where only one
commuting spinor (``twistor variable") was employed, we use eight spinors
satisfying an algebraic condition. They are shown to parametrise the
sphere $S^8$ associated with the lightlike velocity of the particle
\cite{GHS}, \cite{GS}. In Section 3 we extensively
review the twistor formulation of the $D=10$ superparticle of ref.
\cite{GS}, because it serves as the basis of the superstring theory.
The worldline becomes a superspace with one even and eight odd dimensions.
Then the Grassmann coordinates of the target superspace acquire
superpartners with respect to worldline supersymmetry, which are just the
eight spinor variables (twistors). The dynamics of the superparticle is
described by equations which specify the embedding of the worldline
superspace into the target one. The action simply consists of Lagrange
multiplier terms, which impose these embedding conditions on shell. We
also explain the identification of kappa-symmetry of the ordinary
Brink-Schwarz superparticle with local $N=8$ supersymmetry of the
worldline. The coupling of the superparticle to an external Maxwell
superbackground illustrates how the coupling constant (the electric
charge) emerges as an integration constant of the equations of motion.
Geometrically, this coupling means that the pull-back of the Maxwell
one-form onto the worldline becomes pure gauge on shell. In Section 4 we
generalise all these ideas to the heterotic superstring. The worldsheet is
now a heterotic $(p,q)=(8,0)$ superspace.  The Wess-Zumino term of the
superstring involves the two-form superfield from the supergravity
background, which is treated very much like the Maxwell superfield in the
particle case. The difference is that its pull-back onto the worldsheet is
gauge trivial only in the left directions of the heterotic worldsheet.
This time it is the string tension that appears as an integration
constant.

Before proceeding further, a few words about the notation. Throughout the
paper we will have to deal with four different superspaces: the $(1\vert
8)$ or $(2\vert 8)$ worldsheet and its tangent space, and the $(10\vert
16)$ target space and its tangent space. In order to avoid the
proliferation of different types of indices, we adopt the following
notation.  We denote the worldsheet by ${\cal M}$ and the target
superspace by $\un{\cal M}$. We use $M=(m, \mu)$ for coordinate indices
($m$ even and $\mu$ odd) and $A=(a, \alpha)$ for tangent space indices in
${\cal M}$; in $\un{\cal M}$ we use the same conventions but the indices
are underlined.  The coordinates of ${\cal M}$ are denoted $z^M=(x^m,
\theta^\mu)$ and those of $\un{\cal M}$ are denoted $z^{\un
M}=(x^{\un m}, \theta^{\un \mu})$. We shall use the same letters to denote
corresponding quantities in ${\cal M}$ and $\un{\cal M}$.  In the case
that indices are displayed, it will be clear to which space we are
referring from the presence or absence of underlinings; if no indices are
displayed we shall distinguish quantities in $\un{\cal M}$ by underlining
them.  In the case of the superstring in ${\cal M}$ we shall use a
light-cone split for both coordinate and tangent frames, so that $m=(+,-)$
and $a=(+,-)$; it should be clear from the context which type a ``+'' or
``$-$'' index is.

\section{Massless particles, spheres and twistors}

\subsection{The massless $D=10$ particle and the sphere $S^8$}\q{I.1}

The action for a massless particle propagating in a $D=10$ target space is
\be\q{1.1}
S ={1\over 2} \int_{0}^{1} dx^-\; g \;\pl_- x^{\un a}\pl_- x_{\un a} .
\ee
Here $x^-$ denotes the worldline time variable, $x^{\un a}$ are the
coordinates of the $D=10$ target space and $g$ is the worldline metric.
An important kinematical requirement is that the particle velocity of the
particle $\pl_- x^{\un a}$ does not vanish as a vector,
\be\q{no0}
\pl_- x^{\un a}\neq 0 .
\ee
The r\^ole of the metric is to make the action \p{1.1} invariant under
diffeomorphisms of the worldline,
\bea\q{diff}
&&x^- \rightarrow {x'^-}(x^-), \ \ {\pl{x'^-}\over\pl x^-} >0; \\
&&g'(x'^-) = g(x^-) {\pl{x'^-}\over\pl x^-}. \nn
\eea
Note that the diffeomorphisms in \p{diff} preserve the orientation of
time, hence one can regard them as local $SO^\uparrow(1,1)$
transformations.

This action can be rewritten in an equivalent form by passing to the
first-order formalism and introducing an auxiliary vector $u^{\un a}$:
\be\q{1.2}
S = \int dx^-\; \left[p_{\un a}(\pl_- x^{\un a} - u^{\un a}) + {1\over 2}
g u^{\un a} u_{\un a} \right].
\ee
Note that the variation with respect to $g$ makes the vector $u^{\un a}$
lightlike:
\be\q{1.4}
u^{\un a} u_{\un a} = 0.
\ee
This null-vector (which coincides with the velocity of the particle)
transforms as a density under the worldline diffeomorphisms \p{diff},
\be\q{den}
u'(x'^-) = u(x^-) {\pl{x^-}\over\pl x'^-}.
\ee
Further, we can take the time component of this vector to be positive,
since the natural boundary condition $x^0(1)> x^0(0)$ together with
\p{no0}
imply
\be\q{1.5}
\pl_- x^0 > 0 \ \ \rightarrow \ \  u^0 > 0.
\ee

The vector $u^{\un a}$, which satisfies eqs. \p{1.4}, \p{1.5} and is
defined modulo the local $SO^\uparrow(1,1)$ transformations \p{den},
parametrises an eight-dimensional sphere $S^8$. This is easily seen by
choosing the $SO^\uparrow(1,1)$ gauge $u^0=1$ or by dividing the left-hand
side of eq. \p{1.4} by $(u^0)^2$.  Thus one naturally associates a sphere
$S^8$ with the massless particle.

The twistor approach to the particle differs from the standard one in that
the sphere is described by commuting spinor (twistor) variables. The
action
\be\q{act}
S=\int dx^- \;p_{\un a}(\pl_- x^{\un a} - u^{\un a})
\ee
with the constraints \p{1.4}, \p{1.5} on the vector $u^{\un a}$ implied is
the starting point for the transition to a twistor-like particle action.
One possibility, which was employed in \cite{STV}, is to solve the
constraint \p{1.4} in terms of a {\it single} commuting Majorana-Weyl
spinor $\lambda^{\un \alpha}$:
\be\q{twistor}
u^{\un a}=\lambda\gamma^{\un a}\lambda.
\ee
To check that it satisfies \p{1.4} one should use the gamma matrix
identity
\be\q{1.20}
(\gamma^{\un a} )_{\un{(\alpha\beta}} (\gamma_{\un a}
)_{\un{\gamma)\delta}} = 0.
\ee
This identity holds in $D=3,4,6,10$, so the same procedure can be applied
in all these special dimensions. Each time one uses the smallest possible
spinor representations of the corresponding Lorentz groups $SO(1,D-1)$,
which are of real dimensions $2,4,8,16$, respectively.  Note, however,
that the case $D=10$ essentially differs from the other three cases. To
explain this difference we remark that the correspondence between $u^{\un
a}$ and $\lambda^{\un\alpha}$ is not one-to-one: the non-vanishing null
vector $u^{\un a}$ parametrises the spheres $S^{D-2}= S^1, S^2, S^4$ and
$S^8$ in the above dimensions, while the non-vanishing spinor
$\lambda^{\un\alpha}$ parametrises the spheres $S^1, S^3, S^7$ and
$S^{15}$, correspondingly (the vector and the spinor are considered modulo
$SO^\uparrow(1,1)$ scale transformations). The difference between the
vector and spinor degrees of freedom is accounted for by certain gauge
transformations with $0, 1, 3$ and $7$ parameters, respectively.  Now the
point is that while in the lower-dimensional cases $D=3,4$ and $6$ these
gauge transformations form the {\it groups} $Z_2, \; U(1)$ and $SU(2)$ ,
the $7$-parameter gauge transformations in $D=10$ do not correspond to any
{\it group}.  Instead, they correspond to the sphere $S^7$ which is the
fiber in the Hopf fibraion $S^{15}\rightarrow S^8$.  (The lower cases are
the other Hopf fibrations $S^{1}\rightarrow S^1$, $S^{3}\rightarrow S^2$
$S^{7}\rightarrow S^4$ with fibers $Z_2, S^1$ and $S^3$, respectively.
These fibers {\it are} groups, since $S^1=U(1), \; S^3=SU(2)$.) In other
words, the gauge transformations which act on the 16-dimensional
Majorana-Weyl $D=10$ spinor $\lambda^{\un\alpha}$ and leave the vector
$u^{\un a}$ \p{twistor} invariant modulo $SO^\uparrow(1,1)$, do not
constitute any group, their algebra is non-linear and has field-dependent
structure constants. This is an undesired feature of the $D=10$ solution
\p{twistor}, since we want to
construct actions with linear realisations of all symmetries.

Another way to explain why the case $D=10$ is so different from the lower
ones is to note that eqs. \p{1.4} and \p{twistor} are directly related to
the existence of four division algebras (see, for instance, \cite{FM}).
They are equivalent to the defining identity for hypercomplex numbers:
$|ab|^2=|a|^2 |b|^2$. Then the non-linearity and the field-dependence of
the algebra of gauge transformations for the spinor $\lambda$ in
\p{twistor} in the case $D=10$
can be shown to arise from the non-associativity of the octonionic
multiplication (see, for instance, \cite{B}).

Fortunately, there exists another way to describe the sphere $S^8$ in
terms of {\it eight} commuting Majorana-Weyl spinors \cite{GHS}. That
construction is purely real, it makes no use of the notion of hypercomplex
numbers and all the complications mentioned above are avoided.

\subsection{Twistor description of $S^8$}\q{I.2}

The central point in the whole construction of this paper will be to
replace lightlike vectors like $u^{\un a}$ by commuting spinor
(``twistor") variables in a way that maintains all symmetries linearly
realised. Our twistor variables will be defined as follows:
\be\q{1.7}
\lambda_\alpha{}^{\un\alpha}, \ \ \alpha = 1, \ldots,8; \ \ \un\alpha =
1, \ldots, 16,
\ee
where $\alpha$ is an $O(8)$ index \footnote{The choice of an
eight-dimensional representation ($8_s$, $8_c$ or $8_v$) for $\alpha$ is
arbitrary at this point. Later on we shall see that in a special gauge
this $O(8)$ can be identified with the $O(8)$ subgroup of the $D=10$
Lorentz group, and the representation with, for instance, $8_s$.} and
$\un\alpha $ is a Majorana-Weyl $SO(1,9)$ spinor index. The real $8\times
16$ matrix \p{1.7} is defined modulo arbitrary $SO^\uparrow(1,1)\times
O(8)$ transformations
\be\q{gauge}
(\lambda_\alpha{}^{\un\alpha})'=\Omega\omega_\alpha{}^{\beta}
\lambda_\beta{}^{\un\alpha}\;;\;\; \Omega\in SO^\uparrow(1,1), \;
\omega_\alpha{}^\beta \in O(8),
\ee
and is supposed to satisfy the algebraic constraint (invariant under
\p{gauge})
\be\q{1.8}
\lambda_\alpha{}^{\un\alpha} (\gamma^{\un a})_{\un{\alpha\beta}}
\lambda_\beta{}^{\un\beta} = {1\over 8} \delta_{\alpha\beta}
(\lambda_\gamma\gamma^{\un a} \lambda_\gamma),
\ee
or, in matrix notation,
\be
\lambda\gamma^{\un a}\lambda^T = {1\over 8} Tr(\lambda\gamma^{\un
a}\lambda^T) {\bf 1}.
\ee
In addition, the matrix \p{1.7} is required to satisfy the non-degeneracy
condition
\be\q{1.9}
{\rm rank} \parallel \lambda \parallel = 8.
\ee

In order to show how the twistor variables defined above parametrise the
sphere $S^8$ we are going to use light-cone coordinates. Our conventions
are as follows: the ten-dimensional Minkowski metric is
$\eta_{\un{ab}}={\rm diag}(-1,1, \ldots,1)$, the light-cone components of
a vector $p^{\un a}$ are $p^{\pm}=p^0\pm p^9, \;\vec p$, hence $p^{\un a}
p_{\un a}=-p^{+}p^{-}+(\vec p)^2$.  We use the $16\times 16$
$\gamma$-matrix representation where
\bea
\gamma^0=
\left(\begin{array}{cc} 1 & 0 \\ 0 & 1 \end{array}\right), \ \
\gamma^9=
\left(\begin{array}{cc} -1 & 0\\ 0& 1 \end{array}\right), \ \
\gamma^{a_v}=
\left(\begin{array}{cc} 0 & \gamma^{a_v} \\ (\gamma^{a_v})^T & 0
\end{array}\right)
\label{4.3}
\eea
and $(\gamma^{a_v})_{\alpha_s\alpha_c}$ are $8\times 8$ matrices of
$O(8)$, the indices $a_v, \;\alpha_s$ and $\alpha_c$ correspond to the
representations $8_v,
\;8_s$ and $8_c$ of $O(8)$, respectively. Finally, a
Majorana-Weyl spinor $\Theta^{\un\alpha}$ is decomposed into
$(\Theta^{\alpha_s},
\;\Theta^{\alpha_c})$. In this notation the twistor matrix
$\lambda_\alpha{}^{\un\alpha}$ decomposes into two $8\times 8$ matrices,
\be\q{1.10}
\lambda_\alpha{}^{\un\alpha} = (u_{\alpha\alpha_s}, v_{\alpha\alpha_c}),
\ee
in terms of which \p{1.8} reads
\bea
&& uu^T = {1\over 8} Tr(uu^T)\;{\bf 1}, \q{1.11}\\ && u\vec\gamma v^T +
(u\vec\gamma v^T)^T = {1\over 4} Tr(u\vec\gamma v^T) {\bf 1}, \q{1.12}\\
&& vv^T = {1\over 8} Tr(vv^T)\;{\bf 1} . \q{1.13}
\eea

Note that at least one of the matrices $u,v$ is non-degenerate, otherwise
one can derive from \p{1.11} and \p{1.13} that both of them vanish in
contradiction with \p{1.9}. Choosing, for instance,
\be\q{1.14}
\mbox{Chart 1:}\ \ \det u \neq 0,
\ee
we see that \p{1.11} means that $u\in SO^\uparrow (1,1)\times O(8)$.  The
gauge freedom \p{gauge} allows us to choose a gauge in which
\be\q{1.15}
u_{\alpha\alpha_s} = \delta_{\alpha\alpha_s},
\ee
thus identifying the $O(8)$ group of the index $\alpha$ with the subgroup
of the Lorentz group $SO(1,9)$. Note that this gauge breaks Lorentz
invariance.

In the gauge \p{1.15} one can easily solve eq. \p{1.12} for the second
matrix $v$ in terms of an arbitrary $O(8)$ vector $\vec y$:
\be\q{1.16}
v = \vec y\cdot \vec\gamma .
\ee
Then the third equation \p{1.13} is automatically satisfied. The
alternative to \p{1.14} is
\be\q{1.17}
\mbox{Chart 2:}\ \ \det v \neq 0.
\ee
It implies that this time $v\in SO^\uparrow(1,1)\times O(8)$, so it can be
gauged into $\delta_{\alpha\alpha_c}$.  Instead of
\p{1.16} we now find
\be\q{1.18}
u =\vec\gamma^T\cdot \vec z.
\ee
In the overlapping area ( $\det u\neq 0, \; \det v\neq 0$) both
coordinates $\vec y$ and $\vec z$ are non-vanishing. To find the relation
between the two sets of coordinates one can consider the lightlike vector
$\lambda_\alpha\gamma^{\un a}\lambda_\alpha$. Under the gauge group
\p{gauge} it transforms like an $SO^\uparrow(1,1)$ density, so the ratio
of any two of its components is gauge invariant. Computing these ratios on
the first and the second charts one finds
\be
\vec z = \vec y/y^2 ,
\ee
so $\vec y$ and $\vec z$ can be considered as stereographic coordinates on
the sphere $S^8$.

The conclusion from the discussion above is that the twistor variables
defined in \p{1.8}, \p{1.9} describe the sphere $S^8$ modulo
$SO^\uparrow(1,1)\times O(8)$ gauge transformations. The $O(8)$ vectors
$\vec y$ or $\vec z$, corresponding to the choices \p{1.14} or \p{1.17},
are the stereographic coordinates on the two charts needed to cover the
sphere.

We point out that these twistor variables admit a geometric interpretation
as coordinates of a coset space \cite{GHS}, \cite{GS}. The rectangular
matrix $\lambda$
\p{1.7} can be
viewed as one half of a matrix belonging to the spin group $Spin(1,9)$ of
the $D=10$ Lorentz group:
\be\q{Lor}
L_{\un\alpha}{}^{\un\beta} = (\lambda_\alpha{}^{\un\beta},
\phi_{\alpha'}{}^{\un\beta}), \ \ \  L \in Spin(1,9).
\ee
Then conditions \p{1.8}, \p{1.9} appear as that part of the defining
conditions on the $Spin(1,9)$ matrix $L$ \p{Lor}, which concern only the
matrix $\lambda$. The sphere $S^8$ can be thought of as the coset space
\be\label{Spin}
S^8={{\rm Spin}(1,9)\over [SO^\uparrow(1,1)\times O(8)]{\rm
C}\hskip-11pt\times\vec K} .
\ee
Here the denominator is the Borel subgroup of the Lorentz group (the
maximal subgroup of the Lorentz group), including eight ``conformal boost"
transformations $\vec K$. Then the twistor variable matrix $\lambda$ is
that half of $L$ which stays inert under the action of $\vec K$. The other
half, $\phi$, is shown to be either a pure $\vec K$-gauge or is expressed
in terms of $\lambda$ via the remaining $Spin(1,9)$ defining conditions.

\subsection{Twistor particle}\q{I.3}

A key ingredient in the massless particle theory of subsection \ref{I.1}
was the lightlike vector $u^{\un a}$. Now we can replace that vector by a
bilinear combination of the eight spinors (twistor variables) of
subsection \ref{I.2}:
\be\q{1.19}
u^{\un a} = {1\over 8} \lambda_\alpha\gamma^{\un a} \lambda_\alpha.
\ee
To check that it satisfies \p{1.4} one should use \p{1.8} and the gamma
matrix identity \p{1.20}.  The time component of this vector is strictly
positive,
\be\q{1.21}
u^0 = {1\over 8}(Tr(uu^T) + Tr(vv^T)) > 0,
\ee
in accord with \p{1.5}.  The specific twistor combination in
\p{1.19} is $O(8)$ invariant and $SO^\uparrow(1,1)$ covariant,
so it describes $S^8$ modulo $SO^\uparrow(1,1)$ transformations. On the
other hand, in subsection \ref{I.1} we saw that the content of the
lightlike vector $u^{\un a}$ was also $S^8\times SO^\uparrow(1,1)$.  Then
we conclude that the twistor combination \p{1.19} is equivalent to that
vector.

After the discussion above we can propose the following twistor form of
the massless particle action \p{1.2}:
\be\q{1.22}
S = \int dx^-\; \left[ p_{\un a}(\pl_- x^{\un a} - {1\over 8}
\lambda_\alpha\gamma^{\un a} \lambda_\alpha) + p_{\alpha\beta\un a}
\lambda_{\{\alpha}\gamma^{\un
a} \lambda_{\beta\}} \right].
\ee
Here $ p_{\alpha\beta\un a}$ is a new Lagrange multiplier, which imposes
the twistor defining condition \p{1.8} on shell and the symbol
$\{\alpha\beta\}$ denotes the symmetric traceless part
\be
A_{\{\alpha\beta\}}\equiv {1\over 2}(A_{\alpha\beta} + A_{\beta\alpha}
-{1\over 4}\delta_{\alpha\beta}A_{\gamma\gamma}).
\ee

We emphasise that the action \p{1.22} is diffeomorphism invariant without
the help of the worldline metric $g$ of \p{1.2}. Indeed, since the
bilinear combinations of twistors present in \p{1.22} are
$SO^\uparrow(1,1)$ covariant, it is sufficient to identify the
$SO^\uparrow(1,1)$ parameter with $\pl{x'^-}/\pl x^-$. \footnote{Note that
a diffeomorphism transformation is globally not exactly the same as an
$SO^\uparrow(1,1)$ one. This results in an extra boundary term (the
constant invariant length of the particle trajectory) contained in the
twistor variables. In other words, the twistors describe the sphere $S^8$
{\it and} this constant.} This means that the twistor variables take over
the r\^ole of the one-dimensional metric.  The latter reappears in the
process of elimination of the twistors, which leads from \p{1.22} back to
\p{1.2}.

Concluding this subsection, we would like to point out there is an
alternative way to show that the action \p{1.22} describes a massless
particle. It consists in analysing the equations of motion
\bea\q{em1}
&&\delta p_{\un a}: \ \ \pl_- x^{\un a} = {1\over 8}
\lambda_\alpha\gamma^{\un a} \lambda_\alpha, \\
&&\delta x^{\un a}: \ \ \pl_- p_{\un a} = 0, \q{em2}\\
&&\delta\lambda_\alpha{}^{\un \alpha}: \ \ (\gamma^{\un
a}\lambda_\alpha)_{\un\alpha} p_{\un a} = 8 p_{\alpha\beta\un a}
(\gamma^{\un a} \lambda_{\beta})_{\un\alpha}\; , \q{em3}
\eea
following from \p{1.22}, and proving that their gauge invariant content
coincides with that of the equations derived from the standard particle
action \p{1.1}. Using light-cone coordinates and imposing, e.g., the gauge
\p{1.15}, one can show that eq. \p{em3} allows one to solve for $p_{\un
a}$ in terms of the gauge invariant part of the twistors, and thus
identify it with the velocity of the particle (see \p{em1}). At the same
time, the Lagrange multiplier $ p_{\alpha\beta\un a}$ is shown to vanish
on shell, up to gauge degrees of freedom contained in the transformation
\be\q{1.23}
\delta  p_{\alpha\beta\un a}=\Sigma^{\un \alpha}_{\alpha\beta\gamma}
(\gamma_{\un a})_{\un\alpha\un\beta}\lambda_\gamma{}^{\un\beta}.
\ee
Here the parameter $\Sigma^{\un \alpha}_{\alpha\beta\gamma}$ is symmetric
and traceless with respect to its $O(8)$ indices.

\section{Superparticles}

\subsection{The Brink-Schwarz superparticle}\q{II.1}

The Brink-Schwarz superparticle (in a supergravity background) is a
generalisation of the bosonic one
\p{1.1}, where the bosonic $D=10$ target space is replaced by a $(10\vert
16)$ curved target space with tangent group $Spin(1,9)$:
\be
\un{\cal M} = \{x^{\un m}\} \ \ \rightarrow \ \ \un{\cal M} = \{z^{\un M} =
(x^{\un m}, \theta^{\un\mu})\}.
\ee
$\un{\cal M}$ is equipped with preferred frames $E^{\un A} = dz^{\un M}
E_{\un M}{}^{\un A}(z)$ and a Lorentzian connection $\un\Omega$. We take
the constraints on the torsion to be those describing ten-dimensional
supergravity \cite{N}, \cite{Witten}; in particular, we have
\be\q{2.5}
T_{\un{\alpha\beta}}{}^{\un c}= -2i(\gamma^{\un c})_{\un{\alpha\beta}}
\ee
and
\be\q{2.5'}
T_{\un{\alpha\beta}}{}^{\un\gamma} = 0, \ \ T_{\un {a\beta}}{}^{\un c} =
0.
\ee

The superparticle can be viewed as an embedding of the one-dimensional
worldline ${\cal M}$ into the target superspace $\un{\cal M}$. In other
words, the target-superspace coordinates become worldline fields, $z^{\un
M}= z^{\un M}(x^-)$. The object
\be\q{2.1}
E_-{}^{\un A} = \pl_- z^{\un M} E_{\un M}{}^{\un A} (z),
\ee
gives the components of the tangent vector to ${\cal M}$ in the preferred
basis of ${\cal\un M}$.  Then the action of the Brink-Schwarz
superparticle can be written down as the direct analogue of \p{1.1}:
\be\q{2.2}
S ={1\over 2} \int dx^-\; g \;E_-{}^{\un a} E_{-\un a} .
\ee
The geometric meaning of this action is to specify the embedding of ${\cal
M}$ into $\un{\cal M}$ in such a way that the vector $E_-{}^{\un a}$
becomes lightlike on shell.

As before, the action \p{2.2} is worldline diffeomorphism invariant due to
the presence of the metric $g$. In addition, it possesses a somewhat
mysterious gauge symmetry with an anticommuting spinor parameter
$\kappa_{\un\alpha}(x^-)$. Denoting the variation of the target space
coordinates referred to a preferred basis by
\be\q{2.3}
\delta z^{\un A} = \delta z^{\un M} E_{\un M}{}^{\un A}
\ee
the kappa-symmetry transformations can be written as follows \cite{S}:
\be\q{2.4}
\delta {  z}^{\un a} = 0, \ \ \delta {  z}^{\un\alpha} = {
E}_-{}^{\un a} (\gamma_{\un a})^{\un{\alpha\beta}} \kappa_{\un\beta}, \ \
\delta g = 4i g {  E}_-{}^{\un\alpha}\kappa_{\un\alpha} .
\ee
To check the invariance of \p{2.2} one has to use the background
supergravity constraints \p{2.5}, \p{2.5'}.

This symmetry is of crucial importance for the consistency of the
superparticle action. Indeed, take the simplest situation when the target
superspace is flat. In this case the target space supervielbein is given
by
\be
E_{\un m}{}^{\un a} = \delta_{\un m}{}^{\un a}, \ E_{\un m}{}^{\un \alpha}
= 0, \ E_{\un \mu}{}^{\un a} = i(\gamma^{\un a}\un\theta)_{\un\mu}, \
E_{\un \mu}{}^{\un \alpha} = \delta_{\un \mu}{}^{\un \alpha}.
\ee
Then, at first sight the action has the non-linear appearance of an
interacting theory, but this is not the case. The reason is as follows.
The analysis of transformations
\p{2.4} in light-cone coordinates shows that only 8 of the 16 components
of the parameter $\kappa$ are active, because the matrix $E_-{}^{\un a}
\gamma_{\un
a}$ is a projector operator on shell, where the vector $E_-{}^{\un a}$ is
lightlike. These 8 gauge parameters allow one to gauge away half of the
components of the spinor $\theta^{\un\alpha}$. After this has been done
the action is linearised and it is easy to show that the theory is free.
This very important invariance finds its natural explanation as local
$N=8$ supersymmetry of the worldline in the twistor version of the theory.

\subsection{World-line superspace}\q{II.2}

The first step towards a twistor formulation of the superparticle is to
replace the one-dimensional worldline by a $(1\vert 8)$ super-worldline:
\be\q{2.6}
{\cal M} =\{x^-\} \ \ \rightarrow \ \ {\cal M} = \{z^M = (x^-,
\theta^\mu)\}.
\ee
The geometry of ${\cal M}$ is taken to be superconformally flat. Let
$(E_A)$ be a local basis of frames for ${\cal M}$, $E_A=
E_A{}^M\partial_M$, where $E_A{}^M$ is the inverse supervielbein, then the
tangent space group acts by
\be\q{tg}
E_A\rightarrow L_A{}^B E_B , \ {\rm where} \ L_A{}^B=
\left(\matrix{L_-{}^-&L_-{}^\beta\cr
0&L_\alpha{}^\beta\cr}\right) .
\ee
In addition it is supposed that the distribution spanned by $(E_A)$ is
non-integrable, so that the quantity $-i<[E_\alpha,E_\beta]>$ is
non-singular (and positive). Here, ($E^A=dz^M E_M{}^A$) are the basis
one-forms dual to $(E_A)$, and $<,>$ denotes the pairing between forms and
vectors, $<dz^M,
\partial_N>=\delta^M_N$. In other words, the anholonomy coefficient
$C_{\alpha\beta}{}^{-}$ is nonvanishing ($[E_A,
E_B\}=-C_{AB}{}^{{}C}E_C$).

The statement that ${\cal M}$ is superconformally flat (which can be
easily proven given the above structure) means that there exist local
coordinates in which the frames take the standard form of flat superspace,
i.e.
\bea\q{fr}
E_\alpha&=&\partial_\alpha + i\theta_\alpha\partial_- \nn\\
E_-&=&\partial_- . \q{frame}
\eea
When using these basis vectors as derivatives we shall write them as
$D_\alpha$ and $D_-$ respectively. By construction, we have the flat $N=8$
supersymmetry algebra
\be\q{2.8}
\{D_\alpha , D_\beta \} = 2i\delta_{\alpha\beta} D_-, \ \ \
[D_-,D_\alpha ] = 0.
\ee

The class of diffeomorphisms which preserves the form of the frames
\p{frame}
up to tangent space transformations of the form \p{tg} is easily found.
Making a change of coordinates, we have
\be\q{equa}
D_A=D_A z'^M{\partial\over\partial z'^M}.
\ee
Taking the $\alpha$ component of this equation and demanding that
$D_\alpha$ transform into itself, as dictated by the structure of the
tangent space group, we find
\be\q{-}
D_\alpha x'^- -iD_\alpha \theta'^\beta \theta'_\beta = 0.
\ee
This equation shows that the diffeomorphisms in the $\theta$ direction are
expressible in terms of the $x^-$ transformation.  We then have
\be\q{trd}
D_\alpha = L_\alpha{}^\beta D'_\beta,
\ee
where
\be\q{expr}
L_\alpha{}^\beta = D_\alpha \theta'^\beta.
\ee
Differentiating \p{expr} and taking the symmetric part we get the
integrability condition
\be\q{group}
L_\alpha{}^\gamma L_{\beta\gamma} = \delta_{\alpha\beta} L_-{}^-,
\ee
where
\be\q{L-}
L_-{}^- = D_- x'^- - D_- \theta'^\alpha \theta'_\alpha
\ee
Equation \p{group} states that $L_\alpha{}^\beta$ is a matrix belonging to
the group $SO^\uparrow(1,1)\times O(8)$; note also that $L_-{}^-$ is
positive, i.e.  if we regard $x^-$ as a time variable, the allowed
coordinate changes preserve the time orientation. The ``--'' component of
\p{equa} yields no
new constraints on the transformation parameters.

For infinitesimal transformations $\delta z^M = z'^M - z^M$ eq. \p{group}
is linearised and can easily be solved:
\be\q{lambda}
\delta\theta_\alpha = -{i\over 2} D_\alpha \Lambda^-,
\ee
where $\Lambda^-(z)$ is an arbitrary superfield parameter. Inserting this
solution into eq. \p{-}, we can solve for $\delta x^-$ too:
\be\q{xi}
\delta x^- = \Lambda^- - {1\over 2}\theta_\alpha  D_\alpha  \Lambda^-  .
\ee
For the derivatives, we have, with $\delta D_A = D'_A -D_A$,
\be\q{tral}
\delta D_\alpha  = {i\over 2} (D_\alpha D_\beta ) \Lambda^- D_\beta  ,
\ee
\be\q{tr-}
\delta D_- = - (D_-\Lambda^-) D_- + {i\over 2}(D_-D_\alpha \Lambda^-)
D_\alpha ,
\ee
and these transformations are thus explicitly seen to be of the form
\p{tg}. Thus
the group of transformations involves one parameter, $\Lambda^-$, which is
an unconstrained superfield. This group clearly contains $x^-$-space
diffeomorphisms and local supersymmetry transformations.

\subsection{Embedding the worldline in the target superspace}

The motion of the superparticle can be regarded as a specific embedding of
the worldline ${\cal M}$ in the target superspace $\un {\cal M}$.  This
means that the coordinates of ${\cal M}$ are defined as worldline
superfields, $z^{\un M} = z^{\un M} ( z^M)$. To describe this embedding it
is convenient to introduce the matrix
\be\q{2.9}
E_{ A}{}^{\un A} = E_A{}^M(z)\pl_Mz^{\un{M}}(z) E_{\un{M}} {}^{\un A}(\un
z).
\ee
This matrix can be viewed in two ways, as the pull-back of the coframe
$E^{\un A}$ referred to a preferred basis in ${\cal M}$, or as the
push-forward of the frame $E_A$ to $\un {\cal M}$ referred to the
preferred basis of $\un {\cal M}$.

Now we shall specify the way the embedding of ${\cal M}$ in $\un {\cal M}$
is done.  First of all, we remark that the odd-odd part
$E_\alpha{}^{\un\alpha}$ of the matrix \p{2.9} looks very much like the
twistor matrix $\lambda_\alpha{}^{\un\alpha}$. Therefore we impose on
$E_\alpha{}^{\un\alpha}$ the same restriction as the one on
$\lambda_\alpha{}^{\un\alpha}$ in \p{1.8},
\be\q{2.10}
E_{\{\alpha}{}^{\un\alpha}(\gamma^{\un a})_{\un{\alpha\beta}}
E_{\beta\}}{}^{\un\beta} = 0.
\ee
This, together with a non-degeneracy condition similar to \p{1.9},
\be
{\rm rank} \parallel E \parallel = 8,
\ee
allows to identify the lowest-order component of the worldline superfield
$E_\alpha{}^{\un\alpha}$ with the twistor variables:
\be\q{2.11}
\lambda_\alpha{}^{\un\alpha} = E_\alpha{}^{\un\alpha}\vert_{\theta = 0}.
\ee
One can say that the twistor variables emerge as the superpartners of the
target-space odd coordinates with respect to $N=8$ worldline
supersymmetry, \footnote{The possibility to interpret these superpartners
as Lorentz harmonic variables for the case $D=10$ has been pointed out in
\cite{IK}, \cite{GHS}.}
\be
\theta^{\un\alpha}(z) = \theta^{\un\alpha} + \theta_\alpha
\lambda_\alpha{}^{\un\alpha} + \ldots
\ee

As our second embedding condition we demand that the pull-back of a
target-space vector onto a worldline spinor vanish: \footnote{This
equation was proposed in ref. \cite{GHS} (in the case of a flat
background). An analogous equation, but written down in octonionic
notation and in a non-linear and $O(8)$ non-covariant gauge, was studied
in ref. \cite{B}.}
\be\q{2.12}
E_\alpha{}^{\un a} = 0.
\ee
Geometrically, this means that ${\cal M}$ is embedded in $\un {\cal M}$ in
such a way that the odd part of the tangent space to ${\cal M}$ lies
entirely within the odd part of the tangent space to $\un {\cal M}$ at any
point of ${\cal M}$. Thus, at each point in ${\cal M}$ an
eight-dimensional subspace of the odd tangent space to $\un {\cal M}$ is
selected, and the set of such subspaces is parametrised by the coset space
$Spin(1,9)/H$, where $H$ is the Borel subgroup which takes a reference
subspace into itself. As explained earlier, this coset space is the
eight-sphere.

Equation \p{2.12} is a differential equation which requires an
integrability condition. We shall show that, when the constraints on
background supergravity are taken into account, this condition is just
\p{2.10}. First we separate the non-linear
terms in \p{2.12} by writing down the vielbeins $E_{\un{M}}{}^{\un a} =
\delta_{\un{M}}{}^{\un a}
+\mbox{n.l.t.}$ Then eq. \p{2.12} becomes
\be\q{2.13}
D_\alpha x^{\un m} = J_\alpha{}^{\un m} (z^{\un{M}}),
\ee
where we have put all the non-linear terms in the right-hand side of the
equation. Using the flat algebra \p{2.8}, one finds the integrability
condition on the non-linear source:
\be\q{2.14}
D_{\{\alpha} J_{\beta\}}{}^{\un m} = 0.
\ee
To find the covariant version of \p{2.14} we shall use the first of
Cartan's structure equations for $\un {\cal M}$:
\be\q{2.15}
T^{\un A}= d E^{\un A} + E^{\un B} \Omega_{\un B}{}^{\un A}.
\ee
If we pull this back to ${\cal M}$, we can write out the component form as
\be\q{2.16}
D_A E_B{}^{\un C}-(-)^{AB} D_B E_A{}^{\un C} -[E_A,E_B\}^C E_C{}^{\un C}
-\Omega_{AB}{}^{\un C} +(-)^{AB} \Omega_{BA}{}^{\un C}=- T_{AB}{}^{\un C},
\ee
where
\be\q{2.17}
T_{AB}{}^{\un C} = (-)^{A(B+\un B)} E_B{}^{\un B} E_A{}^{\un A}
T_{\un{AB}}{}^{\un C}
\ee
and similarly for $\Omega_{AB}{}^{\un C}$. Putting in \p{2.16} $A=\alpha$,
$B=\beta$, $\un C = \un c$, taking the symmetric traceless part in
$\alpha, \beta$ (as in \p{2.14}) and using once again \p{2.12}, we obtain
\be\q{2.18}
E_{\{\alpha}{}^{\un\alpha}T_{\un{\alpha\beta}}{}^{\un a}
E_{\beta\}}{}^{\un\beta} = 0.
\ee
Comparing this with \p{2.10} we see that they are the same, provided the
background supergravity constraint \p{2.5} holds. Actually, the
compatibility of the integrability condition \p{2.18} with the
sphere-defining condition \p{2.10} on the twistor variables not only
follows from that supergravity constraint, but also implies it (up to
purely conventional parts).  One can show this by decomposing all the
matrices in \p{2.18} into $O(8)$ representations and analysing the
corresponding algebraic relations. This is a manifestation of the
so-called lightlike integrability principle \cite{Witten}, \cite{PURE}.

Another important consequence of the embedding equations \p{2.10},
\p{2.12}
is obtained by putting in \p{2.16} the same values of the indices as
before, but this time taking the trace in $\alpha, \beta$:
\be\q{2.19}
E_-{}^{\un a} = {1\over 8} E_\alpha \gamma^{\un a} E_\alpha.
\ee
Following the arguments of section \ref{I.3}, one sees that this vector is
lightlike,
\be\q{2.20}
E_-{}^{\un a} E_{-{\un a} } = 0.
\ee
The lowest-order component in the $\theta$ expansion of eq. \p{2.20}
coincides with the equation following from the variation with respect to
the worldline metric in the Brink-Schwarz superparticle action
\p{2.2}. As before, the twistor variables describe the sphere $S^8$
associated with this lightlike vector.

The conclusion of this subsection is that by an appropriate embedding of
the $(1\vert 8)$ worldline into the $(10\vert 16)$ target superspace we
have found a natural place for the twistor variables. We also obtained an
important ingredient of the superparticle on-shell dynamics, the lightlike
vector \p{2.20}.

\subsection{The twistor superparticle action}

The action for the twistor superparticle consists of two Lagrange
multiplier terms corresponding to the embedding equations \p{2.10} and
\p{2.12}:
\be\q{2.21}
S= \int {dx^- d^8\theta}\;
\left[iP_{\alpha\un a } E_\alpha{}^{\un a } +
P_{\alpha\beta\un a } E_{\{\alpha}\gamma^{\un a} E_{\beta\}}\right].
\ee
This action is invariant under the special superdiffeomorphisms of
subsection \ref{II.2}. Indeed, in \p{2.21} the spinor worldline indices
transform like the spinor derivatives $D_\alpha$, i.e. homogeneously, see
\p{trd}. It is then easy to find
suitable transformations of the Lagrange multipliers which can compensate
for this, as well as for the transformation of the supervolume.  Once
again, we remark that no worldline supergravity fields are needed to
achieve the diffeomorphism invariance of the twistor action (cf.
subsection \ref{I.3}).

We shall now show that the component content of the action \p{2.21}
reduces to that of the target-space supersymmetric version of the twistor
action \p{1.22}. The transition to the Brink-Schwarz action
\p{2.2} will then follow the pattern of Section 2.

We begin by analysing the dynamical content of eq. \p{2.12}, which follows
from the first term in \p{2.21}. One can show that it produces algebraic
equations for almost all of the components of the superfield $x^{\un
m}(z)$, except for the lowest-order one $x^{\un m}\vert_{\theta=0}$.
Indeed, by dropping all the non-linear terms in \p{2.13} one obtains the
homogeneous form of eq. \p{2.12}, which is $D_\alpha x^{\un m} = 0$. It is
not hard to verify that it sets equal to zero all the components of the
superfield $x^{\un m} (z)$, except for the lowest-order one, which
satisfies the equation $\pl_-x^{\un m}\vert_0 = 0$. The same pattern is
repeated in the inhomogeneous equation \p{2.13}, provided the
integrability condition \p{2.14}, or rather its covariant version \p{2.10}
holds. Thus, we conclude that the only non-trivial component of \p{2.12}
is the lowest-order component of eq. \p{2.19}.

Since all the algebraic component equations contained in \p{2.12} are
introduced in \p{2.21} by Lagrange multipliers (the components of
$P_{\alpha\un a}$), one can drop them in the action. Thus the only
surviving component of the first term in $S$ is
\be\q{6}
\int dx^-\; p_{\un a} \left({\cal E}_-{}^{\un a} - {1\over 8}
{\lambda}_\alpha \gamma^{\un a} {\lambda}_\alpha \right).
\ee
Here we have used the notation \p{2.11} and have denoted the only relevant
component of the Lagrange multiplier by $p_{\un a} = (D^7)_\alpha
P_{\alpha\un a}\vert_0$ and the lowest-order component of $E_-{}^{\un a}$
by ${\cal E}_-{}^{\un a} = E_-{}^{\un a}\vert_0$.

Next we discuss the second term in $S$, which introduces the
$S^8$-defining constraint \p{2.10}. The analysis of this equation goes
along the lines of subsection \ref{I.2}. Using light-cone notation, one
can decompose the matrix $E_\alpha{}^{\un\alpha}$ into two $8\times 8$
matrices
\be
E_\alpha{}^{\un\alpha} = (E_{\alpha\alpha_s}, F_{\alpha\alpha_c}),
\ee
which satisfy equations similar to \p{1.11}-\p{1.13}:
\bea
&& EE^T = {1\over 8} Tr(EE^T)\;{\bf 1}, \q{sph1}\\ && E\vec\gamma F^T +
(E\vec\gamma F^T)^T = {1\over 4} Tr(E\vec\gamma F^T) {\bf 1}, \q{sph2}\\
&& FF^T = {1\over 8} Tr(FF^T)\;{\bf 1} . \q{sph3}
\eea

To understand the implications of eq. \p{sph1} we note the following.
Suppose that the matrix $E_{\alpha\alpha_s}$ is non-degenerate (this
corresponds to choosing one of the two charts on the sphere $S^8$). This
matrix $E_{\alpha\alpha_s}$ transforms homogeneously under the finite
diffeomorphisms of subsection \ref{II.2},
\be\q{trn}
E_{\alpha\alpha_s} = (D_\alpha\theta'_\beta) E'_{\beta\alpha_s}.
\ee
In \p{group} we have seen that the parameter matrix in \p{trn} satisfies
the same condition as $E$ in \p{sph1}. On the other hand, the matrix
$D\theta$ satisfies the differential constraint
\be\q{dc}
D_{\{\alpha} (D_{\beta\}}\theta'_\gamma) = 0.
\ee
It turns out that the matrix $E$ satisfies the same type of constraint.
This can be derived from \p{2.16}, \p{2.5'} and \p{2.12}. Using the
non-degeneracy of $E_{\alpha\alpha_s}$, one is able to fix the light-cone
gauge
\be\q{3}
E_{\alpha\alpha_s} = \delta_{\alpha\alpha_s} \ \ \rightarrow \ \
\un\theta^{\alpha_s}(z) = \theta^{\alpha_s}.
\ee
This gauge fixes most of the worldline superdiffeomorphisms and identifies
the $O(8)$ automorphism group of worldline supersymmetry with the $O(8)$
subgroup of the tangent-space $D=10$ Lorentz group, thus breaking Lorentz
invariance. Note that in the action we shall not need to fix that gauge,
thus preserving all the gauge symmetry of the worldline, including its
local $N=8$ supersymmetry. For us the essential conclusion from the
discussion above will be that eq. \p{sph1} is purely auxiliary.

Equation \p{sph2} allows one to solve for the matrix $F$ in terms of an
$O(8)$ vector-valued superfield $Y_{a_v}(z)$:
\be\q{4}
F = \vec Y\cdot \vec\gamma.
\ee
This equation is auxiliary too. To see that one should look at it in the
gauge \p{3} and in the linearised limit with respect to background
supergravity:
\be\q{5}
D_\alpha \un\theta_{\alpha_c} = (\gamma_{a_v})_{\alpha\alpha_c} Y_{a_v}.
\ee
It is clear that part of $\un\theta^{\alpha_c}$ is eliminated, and the
rest defines the components of $Y$. The essential point is that the only
component of the left-hand side of eq. \p{5}, which could possibly give
rise to an equation of motion, is absorbed by a component of $Y$:
\be
8i \pl_-\un\theta_{\alpha_c} \vert_0 = (\gamma_{a_v})_{\alpha\alpha_c}
D_{\alpha} Y_{a_v}\vert_0 .
\ee

Finally, the third equation \p{sph3} is not independent, it is solved by
\p{4}. As we explained in subsection \ref{I.2}, the whole argument above
can be turned around by choosing the other chart on $S^8$, where $\det
F\neq 0$.

So, we have seen that equations \p{sph1}-\p{sph3} are purely auxiliary.
They put algebraic restrictions on the components. In particular, the
twistor variables \p{2.11} satisfy their defining condition \p{1.8}.
Therefore in the component action we can keep only the Lagrange multiplier
term imposing that constraint. Putting this together with \p{6}, we obtain
the component form of the action \p{2.21}:
\be\q{compact}
S= \int dx^-\; \left[p_{\un a} \left({\cal E}_-{}^{\un a} - {1\over 8}
{\lambda}_\alpha \gamma^{\un a} {\lambda}_\alpha \right) +
p_{\alpha\beta\un a}
\lambda_{\{\alpha}\gamma^{\un
a} \lambda_{\beta\}} \right].
\ee
We see that this is the target-space supersymmetrised version of the
twistor action \p{1.22}. To obtain the Brink-Schwarz action \p{2.2}, we
have to vary with respect to the Lagrange multiplier $p_{\alpha\beta\un
a}$ and $p_{\un a}$, and then eliminate the twistor variables replacing
them by the lightlike vector $u^{\un a}=1/8{\lambda}_\alpha \gamma^{\un a}
{\lambda}_\alpha$.  In order to account for the lightlike condition on
that vector, we need to introduced a new Lagrange multiplier, the
worldline metric $g$:
\be\q{BS}
S = {1\over 2}\int dx^-\; g\; {\cal E}_-{}^{\un a}{\cal E}_{-\un a}.
\ee

In the process of obtaining the component action \p{compact} the Lagrange
multiplier superfields in \p{2.21} were almost entirely eliminated. To a
large extent this is due to a powerful abelian gauge invariance:
\be\q{sigma}
\delta P_{\alpha\un a} = (D_\beta\delta_{\un a}{}^{\un b}+
\Omega_{\beta{\un a}}{}^{\un b} -{1\over 2}E_\beta^{\un c}
T_{\un c{\un a}}{}^{{}{}\un b})\Sigma_{\alpha\beta\un b}, \ \
\delta P_{\alpha\beta\un
a} = \Sigma_{\alpha\beta\un a},
\ee
where $\Sigma_{\alpha\beta\un a}$ is an arbitrary parameter which is
completely symmetric and traceless with respect to its worldline indices.
The necessary and sufficient condition for this transformation to be a
symmetry of the superspace superparticle action is that the supergravity
constraint \p{2.5} should hold (see \cite{GS} for more details). Note that
the transformations \p{sigma} contain the component gauge transformation
\p{1.23}, the r\^ole of which was explained in subsection
\ref{I.2}.

Concluding this subsection, we would like to compare our approach to the
one recently proposed by Tonin \cite{TON}. Although he only discusses the
superstring, part of his action can be applied to the superparticle as
well. He writes down an equation of motion of the type of eq. \p{2.12}
\footnote{Note that
eq. \p{2.12} already appeared in ref. \cite{GHS} (see also \cite{B}).} and
derives from it the analogue of
\p{2.10}, which defines his space of twistor variables. The principal
difference is that he uses complex worldsheet Grassmann variables (his
index $\alpha$ is a $U(4)$ one, and not an $O(8)$ one). As a result, his
twistor variables parametrise a 20-dimensional coset space,
\be\label{Spin1}
{{\rm Spin}(1,9)\over [SO^\uparrow(1,1)\times U(4)]{\rm
C}\hskip-11pt\times \vec K} ,
\ee
which is called the space of projective pure spinors (see \cite{FR},
\cite{PURE}, \cite{GHS}). As we saw
earlier, the massless particle in ten dimensions is naturally associated
with the sphere $S^8$, which is represented by the coset \p{Spin}, so pure
spinors are not necessary for our purposes.  \footnote{At the same time,
pure spinors seem to have an advantage from the point of view a lightlike
integrability interpretation of the constraints of $D=10$ supergravity,
see \cite{PURE}. Future development of the subject may show if one would
need pure spinors for some applications. If so, switching from $S^8$ to
the space of pure spinors is easy, since the space \p{Spin1} is locally
isomorphic to the tensor product of $S^8$ with the compact coset
$O(8)/U(4)$. Then one could add a set of new $O(8)$ harmonic variables to
our twistor variables.} We shall comment on further substantial
differences between the approach of Tonin and ours in the section devoted
to the superstring.

\subsection{Kappa-symmetry from worldline supersymmetry}

The superspace action $S$ \p{2.21} is invariant under the worldline
superdiffeomorphisms of subsection \ref{II.2}. In the process of deriving
the component form of $S$ we have just eliminated a number of auxiliary
fields. Therefore we can be sure that the action \p{BS} has retained its
original symmetry. In particular, it must be invariant under local $N=8$
supersymmetry, which is part of the diffeomorphism supergroup.  The
corresponding transformations of the dynamical variables $z^{\un
M}\vert_0$ can be obtained as follows:
\be\q{sus'}
\delta z^{\un A} = \delta z^{\un M} E_{\un M}{}^{\un
A}\vert_0 = \epsilon_{\alpha} D_{\alpha}z^{\un M} E_{\un M}{}^{\un
A}\vert_0 = \epsilon_{\alpha} {\cal E}_{\alpha}{}^{\un A}.
\ee
The local supersymmetry parameter $\epsilon^{\alpha}(x^-)$ is the
lowest-order term in the superdiffeomorphism transformation
$\delta\theta^{\alpha} =
\epsilon^{\alpha}(x^-) + \ldots$. It is precisely this parameter that
can be used to gauge away half of the Grassmann coordinates of target
superspace, $\un\theta^{\alpha_s}(x^-) =
\un\theta^{\alpha_s}(z)\vert_{\theta=0}$ (cf. the gauge \p{3}).
Using \p{2.12}, one finds from \p{sus'}:
\be\q{susy}
\delta z^{\un a} = 0, \ \ \delta z^{\un\alpha} =
\epsilon_{\alpha} {\lambda}_{\alpha}{}^{\un\alpha}.
\ee
These transformations are Lorentz covariant due to the presence of the
twistor variables ${\lambda}_{\alpha}{}^{\un\alpha}$, which ``push
forward" the worldline $O(8)$ spinor parameter $\epsilon_\alpha$ to a
target-space Lorentz spinor. However, to obtain \p{BS} we have eliminated
the twistors from \p{compact}, so we have to do the same in
\p{susy}. In \p{compact} this was easy, because the twistors appeared
only in specific quadratic combinations. In
\p{susy} they appear linearly, so we need to express them in terms
of the dynamical variables. This can be done using the results of
subsection \ref{I.2} and the ``--" and transverse projections of the
equation ${\cal E}_-{}^{\un a} - {1/8} {\lambda}_\alpha\gamma^{\un a}
{\lambda}_\alpha=0$ following from the action \p{compact}. The result is
\be\q{solution}
\lambda_\alpha{}^{\un\alpha} = (u_{\alpha\alpha_s}, v_{\alpha\alpha_c}): \
\ \ {u}_{\alpha\alpha_s} = \delta_{\alpha\alpha_s} \sqrt{{\cal E}^-_-\over
2}, \
\ \ {v}_{\alpha\alpha_c} = {1\over \sqrt{2{\cal E}^-_-}}\vec{\cal
E}_- \cdot \vec\gamma_{\alpha\alpha_c} .
\ee
Here we have gauged away the $O(8)$ part of the matrix ${u}$, thus
actually breaking Lorentz invariance by identifying the worldline group
$O(8)$ with the subgroup of the target-space Lorentz group.  \footnote{A
gauge of the type \p{3} would be too strong here, because it fixes the
$N=8$ worldline local supersymmetry.} Further, in deriving \p{solution} we
had to make the non-covariant assumption
\be\q{ass}
{\cal E}^-_- > 0.
\ee
This is possible, since the lightlike vector ${\cal E}_-{}^{\un a}$ has a
twistor origin, so ${\cal E}^-_- {\cal E}^+_- = (\vec{\cal E})^2 \geq 0$
and ${\cal E}^-_- + {\cal E}^+_- = {\cal E}^0_- > 0$. This means that at
least one of the components ${\cal E}^-_-$ or ${\cal E}^+_- $ is strictly
positive. Actually, the assumption \p{ass} corresponds to one of the
charts on $S^8$; the other choice ${\cal E}^+_- > 0$ corresponds to the
second chart and results in an alternative form of \p{solution}. So,
putting \p{solution} in
\p{susy}, we obtain
\be\q{sus}
\delta z^{\un a} = 0, \ \ \delta z^{\alpha_s} =
\sqrt{{\cal E}^-_-\over 2}\epsilon^{\alpha_s}, \ \
\delta z^{\alpha_c} = {1\over \sqrt{2{\cal E}^-_-}}\vec{\cal
E}_- \cdot (\vec\gamma\epsilon)^{\alpha_c}.
\ee
Finally, using the background torsion constraints, one can verify that the
action \p{BS} is invariant under \p{sus}, provided the metric transforms
as follows \footnote{In fact, with the assumption
\p{ass} one can show that the metric in \p{BS} originates from
a component of the Lagrange multiplier $p_{\un a}$ in \p{compact}, $g =
p^- /{\cal E}^-_-$. Then this implies the transformation law \p{g}.}
\be\q{g}
\delta g = -{2\sqrt{2}i\over \sqrt{{\cal E}^-_-}} g
\epsilon^{\alpha_s} {\cal
E}_-{}^{\alpha_s}.
\ee

So, we see two distinct possibilities. The first one is the twistor
action, where the local $N=8$ supersymmetry transformations
\p{susy} are Lorentz covariant. The other one is the action \p{BS}, which
only involves the standard fields of the Brink-Schwarz superparticle, but
where supersymmetry is non-covariant. One can say that in the twistor
version Lorentz covariance is due to the $O(8)$ gauge degrees of freedom
present in the twistor variables. Remarkably, by adding new gauge degrees
of freedom (this time fermionic ones), one can restore the Lorentz
covariance of the supersymmetry transformations \p{sus}, \p{g}. Namely,
one can replace the 8 parameters $\epsilon^{\alpha}$ by a twistor
projection of a full 16-component $D=10$ spinor parameter
$\kappa_{\un\alpha}$:
\be\q{k'}
\epsilon_\alpha = {\lambda}_\alpha{}^{\un\alpha} \kappa_{\un\alpha},
\ee
or, using the explicit solution \p{solution},
\be\q{k}
\epsilon_{\alpha_s} = \sqrt{{\cal E}^-_-\over 2}\kappa_{\alpha_s} +
{1\over \sqrt{2{\cal E}^-_-}} \kappa_{\alpha_c} (\vec\gamma \cdot
\vec{\cal
E}_-)_{\alpha_c\alpha_s} .
\ee
Putting this in the transformations laws \p{sus} and \p{g}, one sees that
they take the Lorentz-covariant form \p{2.4} of kappa symmetry (up to
trivial terms proportional to the equations of motion).  The conclusion is
that both the twistor and the Brink-Schwarz superparticle actions have
local $N=8$ worldline supersymmetry, but its Lorentz-covariant
realisations are different.

\subsection{Maxwell coupling}\q{II.6}

A very interesting feature of the twistor superparticle is revealed when
coupling it to a background Maxwell superfield $A_{\un M}(\un z)$. The
latter undergoes abelian gauge transformations
\be\q{m1}
\delta A_{\un M}(\un z) = \pl_{\un M} a (\un z)
\ee
and satisfies the constraints
\be\q{m2}
F_{\un{\alpha\beta}} = 0, \ \ \ F_{\un\alpha\un b} = (\gamma_{\un
b}F)_{\un \alpha}
\ee
(in $D=10$ these constraints are on-shell). Here
\be\q{m3}
F_{\un A\un B} = (-)^{{\un A}({\un B}+{\un N})} E_{\un B}{}^{\un N} E_{\un
A}{}^{\un M}F_{\un M\un N}
\ee
and $F_{\un M\un N} = \pl_{\un M} A_{\un N} - (-)^{\un M\un N} \pl_{\un N}
A_{\un M}$ is the field-strength tensor.

Consider now the pull-back of $A_{\un M}$ to the worldline
\be\q{m4}
A_M = \pl_M z^{\un M} A_{\un M}.
\ee
It undergoes the gauge transformations
\be\q{m5}
\delta A_M = \pl_M z^{\un M} \pl_{\un M} a (\un z) = \pl_M  a (z).
\ee
One can show that the pull-back of the field-strength $F_{AB} =
(-)^{A(B+{\un B})} E_B{}^{\un B} E_A{}^{\un A}F_{\un A\un B}$ vanishes as
a consequence of the constraints \p{m2} and the superparticle equations of
motion. Indeed, using \p{2.12} and \p{m2} one finds
\be
F_{\alpha\beta} = E_\alpha{}^{\un\alpha} E_\beta{}^{\un \beta} F_{\un{
\beta\alpha}} = 0.
\ee
Further, from \p{2.12}, \p{2.19}, \p{2.10} and \p{m2} one obtains
\be
F_{\alpha -} = E_\alpha{}^{\un\alpha} E_-{}^{\un b} F_{\un\alpha \un b} =
E_\alpha{}^{\un\alpha} {1\over 8}(E_\beta\gamma^{\un b}E_\beta)
(\gamma_{\un b}F)_{\un \alpha} = 0.
\ee
So,
\be\q{m6}
F_{AB} = 0 \ \ \leftrightarrow \ \ F_{MN} = 0.
\ee

Now we can write down the superparticle-Maxwell coupling term
\be\q{m7}
S_{Maxw} = \int dx^-d^8\theta \; (A_M - \pl_M Q) P^M.
\ee
Here $P^M$ and $Q$ are Lagrange multipliers. The equation following from
the variation with respect to $P^M$ is
\be\q{m8}
A_M = \pl_M Q.
\ee
Comparing it with \p{m5}, one can say that on shell the pull-back $A_M$
becomes a pure gauge. There is an integrability condition for equation
\p{m8}, and it is just \p{m6}. In fact, we can turn this argument the
other way around. Demanding integrability for \p{m8}, we arrive at \p{m6}.
{}From there, using the properties of the twistor variables, we can derive
the super-Maxwell constraints \p{m2}. This is another example of lightlike
integrability.

The implications of the equation of motion for $Q$ are very interesting.
It reads \be\q{m9}
\pl_M P^M = \pl_- P^- + \pl_\mu P^\mu = 0.
\ee
It is easy to see that the general solution of eq. \p{m9} is
\be\q{m10}
P^M = \pl_N \Sigma^{NM} + \theta^8 \delta_-{}^M\;e, \ \ \ e = {\rm const}.
\ee
Here $\Sigma^{MN}(z)$ is an arbitrary worldline superfield with graded
antisymmetry in $M,N$ and $e$ is an integration constant. The origin of
this constant term has to do with the non-trivial cohomology of
superspace.  Indeed, the second term in \p{m9} contains a Grassmann
derivative, so in its $\theta$ expansion the highest-order term $\theta^8$
is missing. This yields the constraint $\pl_- e = 0$ on the highest-order
component of $P^-$.  The $\Sigma$ term in \p{m10} corresponds to a gauge
symmetry of the action \p{m7}, which follows from the constraint \p{m6}.
Using this gauge freedom one can fix an on-shell gauge for \p{m9}, where
it reduces to just one constant component:
\be\q{m11}
P^- = e\theta^8, \ \ \ P^\mu = 0.
\ee
In the process of fixing that gauge one does not have to use parameters
with time derivatives, so the gauge is globally achievable and can be
inserted in the action \p{m7}. The result is the usual component
superparticle-Maxwell coupling:
\be
S_{ Maxw} = \int dx^- d^8\theta \; \theta^8\; e A_-(z) = \int dx^- \;
e\;\pl_-z^{\un M} A_{\un M}(\un z)\vert_0.
\ee
We see that if the constant $e$ is non-vanishing, it has the meaning of an
electric charge. In the original superspace action \p{m7} the electric
charge was not manifestly present, it was represented by a worldline field
(the last component of the Lagrange multiplier $P^-$). Only on shell did
it become a constant and acquire its standard physical meaning.

Finally we remark that the geometric meaning of the Lagrange multiplier
$Q$ is that of a Kaluza-Klein coordinate.  Indeed, under the Maxwell gauge
transformation \p{m5} it is shifted by $\delta Q= a ({\underline z})$ and
neither the fields nor the gauge parameters depend on it.

\section{Heterotic superstrings}

\subsection{The Green-Schwarz superstring}

The principal difference between the Brink-Schwarz superparticle and the
Green-Schwarz superstring is that the one-dimensional worldline of the
former becomes a two-dimensional worldsheet for the latter: \be\q{3.1}
{\cal M} = \{x^-\} \ \ \rightarrow \ \ {\cal M} = \{x^{m} = (x^+,x^-)\} .
\ee
This allows one to write down an action which incorporates a sigma-model
term similar to \p{2.2}, and a new, Wess-Zumino term involving the
two-form of background supergravity:
\be\q{3.2}
S = \int d^2x\left(e^\Phi \sqrt{-g} g^{mn} E_m{}^{\un a} E_{n\un a} +
\pl_-z^{\un N}\pl_+z^{\un M} B_{\un{MN}} \right).
\ee
Here $g^{mn}$ is the metric of the two-dimensional worldsheet, $E_m{}^{\un
a} = \pl_mz^{\un M} E_{\un M}{}^{\un a}$ is a vielbein pull-back,
$\Phi(\un z)$ is the $D=10$ supergravity dilaton superfield and
$B_{\un{NM}}(\un z)$ is the two-form superfield. The field-strength of the
latter, the three-form
\be
H_{\un{ABC}} =(-)^{({\un A}+{\un B})({\un K}+{\un C})+{\un A}({\un B}+{\un
N})} E_{\un C}{}^{\un K} E_{\un B}{}^{\un N} E_{\un A}{}^{\un M} (\pl_{\un
M}B_{\un{NK}} + {\rm graded cycle}),
\ee
is supposed to satisfy the supergravity constraints
\be\q{3.3}
H_{\un{\alpha\beta\gamma}} = 0
\ee
and
\be\q{3.4}
H_{\un{a\beta\gamma}}= -2ie^\Phi (\gamma_{\un a})_{\un{\beta\gamma}}.
\ee

Note that the sigma-model term in \p{3.2} needs the two-dimensional metric
for diffeomorphism invariance. Because of the specific, Weyl invariant
combination $\sqrt{-g} g^{mn}$ only two of the three components of that
metric actually appear in \p{3.2}. The variation with respect to them
gives rise to the two Virasoro constraints
\be\q{Virasoro}
E_-{}^{\un a} E_{-\un a} = 0, \ \ E_+{}^{\un a} E_{+\un a}=0
\ee
(in the gauge $g^{mn} = \eta^{mn}$). Note that the metric does not appear
in the Wess-Zumino term, which is a topological invariant. The kappa
symmetry \p{2.4} of the superparticle can be generalised to the case of
the superstring action
\p{3.2}, taking
into account the new background constraints \p{3.3}, \p{3.4}. Once again,
it plays the crucial r\^ole of rendering the theory free when the
background is flat. By supersymmetrising the worldsheet and thus
introducing twistor variables, we shall be able to replace kappa-symmetry
by local supersymmetry of the worldsheet. The first of the Virasoro
constraints \p{Virasoro} will be obtained via the twistor mechanism, while
for the second one we shall still need one component of the worldsheet
metric.

\subsection{Two-dimensional super-worldsheet}

Our first step is to replace the worldsheet \p{3.1} by a $(2\vert 8)$
super-worldsheet:
\be\q{3.5}
{\cal M} = \{x^m\} \ \ \rightarrow \ \ {\cal M} = \{z^{M} = (x^m,
\theta^\mu)\} .
\ee
It is natural to choose the supersymmetry of the worldsheet to be of the
heterotic type $N=(8,0)$, i.e.  the algebra of the flat covariant
derivatives is given by
\be\q{alder}
\{D_\mu, D_\nu\}=2i\delta_{\mu\nu}\partial_-, \;\;\;\;
D_\mu=\partial_\mu+i\theta_\mu\partial_-.
\ee
So, supersymmetry only affects the left-handed even coordinate $x^-$, as
well as the odd ones $\theta^\mu$, but does not affect the right-handed
coordinate $x^+$.

The reason for this choice is as follows. The target and worldsheet
manifolds are closely related to each other since one of them is embedded
into the other. Given a point on the worldsheet ${\cal M}$ we can always
choose a frame in the target space $\un{\cal M}$ with two axes in the
tangent plane to the worldsheet. In this frame the $N=1 \;D=10$
target-space supersymmetry algebra $\{D_{\un\alpha}, D_{\un\beta}\}=
2i(\gamma^{\un a})_{\un\alpha\un\beta} \partial_{\un a}$ reads
\be\q{10alg}
\{D_{\alpha_s}, D_{\beta_s}\}=2i\delta_{\alpha_s\beta_s}\partial_-, \;\;
\{D_{\alpha_c}, D_{\beta_c}\}=2i\delta_{\alpha_c\beta_c}\partial_+, \;\;
\{D_{\alpha_s}, D_{\beta_c}\}=2i\gamma^{a_v}_{\alpha_s\beta_s}\partial_{a_v},
\ee
where $\alpha_s, \;\alpha_c$ and $a_v$ are indices of the $8_s, \;8_c$ and
$8_v$ representations of $O(8)$, correspondingly. In general, the
supersymmetry transformations in the target space induce some
supersymmetry on the worldsheet. From \p{10alg} it is perfectly clear that
only one half of the 16 generators $D_{\un\alpha}$, either $D_{\alpha_s}$
or $D_{\alpha_c}$ (but not both!) do not generate translations in the
transverse direction, which are orthogonal to the worldsheet. So, the
supersymmetry induced on the worldsheet is of the type $(8,0)$.  It should
be emphasized that the heterotic nature of the worldsheet is a direct
consequence of the chiral structure of $N=1\;D=10$ supersymmetry.  There
are two inequivalent 16-dimensional Majorana-Weyl spinors in $D=10$, ${\bf
16}$ and ${\bf 16'}$, and $N=1$ supersymmetry involves only one of them.
Since ${\bf 16}$ and ${\bf 16'}$ are related by the time reflection $T$
one can say that $T$ is broken in a maximal way.  The same is true for the
worldsheet $(8,0)$ supersymmetry \p{alder}.

Another argument in favour of the above choice of the worldsheet is
related to the $\kappa$-symmetry of the ordinary Green-Schwarz
superstring. This symmetry effectively involves $8$ anticommuting gauge
parameters, which we are going to interpret as local worldsheet
supersymmetry later on.

The geometry of ${\cal M}$ is defined by two requirements. First, the
Grassmann worldsheet derivative should transform homogeneously, which
means that the tangent space group action is given by (cf. \p{tg})
\be\q{3.6}
L_A{}^B=
\left(\matrix{L_+{}^+&L_+{}^-&L_+{}^\beta\cr
0&L_-{}^-&L_-{}^\beta\cr 0&0&L_\alpha{}^\beta\cr}\right).
\ee
Second, there should be no objects invariant under \p{3.6} and the
worldsheet superdiffeomorphism group. These requirements are sufficient
for the existence of local coordinates in which the frames take the
standard form of flat superspace (cf. \p{fr}), i.e.
\bea
E_\alpha &=& \partial_\alpha + i\theta_\alpha\partial_- \nn\\ E_- &=&
\partial_- \q{3.7}\\
E_+&=&\partial_+ . \nn
\eea
This form is preserved up to local tangent space rotations by a group of
generalised superconformal coordinate transformations. Indeed, repeating
the argument of subsection
\ref{II.2}, we see that $E_\alpha$
and $E_-$ are preserved by left-handed superdiffeomorphisms similar to
\p{lambda}-\p{tr-}. Clearly,
these involve an unconstrained left-handed even diffeomorphism parameter
$\delta x^- = \lambda(x^+,x^-)$. However, the right-handed transformations
compatible with the frames \p{3.7} have to be restricted to conformal
ones, i.e. $\delta x^+ = \lambda(x^+)$. If we write down a superstring
action which only has this symmetry, we are going to miss the second of
the Virasoro constraints \p{Virasoro}. Therefore we shall need a slightly
bigger group for our purposes. This can be achieved by making coordinate
and tangent space transformations on the above frames to bring them to the
form
\bea
E_\alpha &=& \partial_\alpha + i\theta_\alpha\partial_- + E_\alpha{}^+
\partial_+ \nn\\
E_-&=&\partial_- - {i\over 8}D_\alpha E_\alpha{}^+\pl_+ \q{3.8}\\
E_+&=&\partial_+ . \nn
\eea
When using these basis vectors as derivatives we shall write them as
$D_A$. By construction, they satisfy the flat algebra \p{2.8}, which
implies the constraint
\be\q{3.9}
D_{\{\alpha} E_{\beta\}}{}^+ = 0.
\ee

Once more, we require that the allowed worldsheet superdiffeomorphisms
preserve the form of the frames \p{3.8} up to tangent space
transformations of the form \p{3.6}, cf. \p{equa}. The resulting
constraints are \p{-} and
\be\q{3.10}
D_\alpha x'^+ - D_\alpha \theta'^\beta E'_\beta{}^+ = 0.
\ee
The new equation \p{3.10} gives the transformation rule of $E_\alpha{}^+$.

In infinitesimal form, the transformation laws of the derivatives
$D_\alpha, D_-$ are the same as in the particle case, see
\p{lambda}-\p{tr-}. To these we have to add the shift of the new
worldsheet coordinate,
\be\q{3.11}
\delta x^+ = \Lambda^+(z)
\ee
with an arbitrary superfield parameter, the transformation law of
$E_\alpha{}^+$,
\be\q{3.12}
\delta E_\alpha{}^+ = D_\alpha \Lambda^+ + {i\over 2} D_\alpha D_\beta
\Lambda^- E_\beta{}^+
\ee
and that of the derivative $D_+$,
\bea
\delta D_+ &=& -[D_+\Lambda^+ +{i\over 2} (D_+D_\alpha\Lambda^-)
E_\alpha{}^+ + {i\over 8} (D_+\Lambda^-)D_\beta E_\beta{}^+] D_+ \nn\\ &-&
(D_+\Lambda^-) D_- + {i\over 2}(D_+D_\alpha \Lambda^-) D_\alpha. \q{DDD}
\eea
Clearly, $D_A$ transform in accordance with \p{3.6}.

The group of superdiffeomorphisms above involves two parameters,
$\Lambda^+,
\Lambda^-$, both of which
are unconstrained superfields. This group clearly contains $x$-space
diffeomorphisms and local supersymmetry transformations. It should be
emphasised that this group is larger than the $(8,0)$ superconformal
group. The latter is obtained by setting $E_\a{}^+$ equal to zero and
requiring that the derivative $D_+$ transfroms into itself, i.e.  by
restricting the tangent space group such that $L_+{}^-=L_+{}^\alpha=0$.
{}From \p{3.12} and
\p{DDD} we find that these restrictions lead to the holomorphic parameters
$\pl_+\Lambda^- =0 \ \rightarrow \ \Lambda^- = \Lambda^-(x^-,\theta)$,
$D_\alpha
\Lambda^+= D_-\Lambda^+ = 0 \ \rightarrow \ \Lambda^+ = \Lambda^+(x^+)$
of the superconformal group.  The superconformal symmetry is
characteristic for all actions with dimensionless coupling constants.
However, our action has an additional Chern-Simons structure which
explains the larger group.

Note also that if only $E_\a{}^+$ is set equal to zero, but the derivative
$D_+$ is not required to transform homogeneously, the parameter
$\Lambda^-$ remains unconstrained while the parameter $\Lambda^+$ is
restricted as before, $\Lambda^+ = \Lambda^+(x^+)$. This defines an
intermediate generalised superconformal group.  The gauge $E_\alpha{}^+ =
0$ is not globally possible and is thus not allowed in the action.
Nevertheless, there exists a slightly weaker Wess-Zumino gauge for
$E_\alpha{}^+$:
\be\q{3.13}
E_\alpha{}^+\vert_{WZ} = i\theta_\alpha g_{--}(x) .
\ee
In this gauge the covariant derivative $D_-$ (see \p{3.8}) becomes
\be\q{3.14}
D_- = \pl_- + g_{--}\pl_+ .
\ee
To achieve that gauge one uses only those parameters in the decomposition
of $\Lambda^+$, which enter \p{3.12} without space-time derivatives.  The
remaining gauge field $g_{--}(x)$ is the component of the two-dimensional
metric corresponding to the purely bosonic right diffeomorphisms $\delta
x^+ = \Lambda^+\vert_{\theta=0}$. In fact, the necessity to keep the gauge
field $g_{--}(x)$ (which is responsible for the second Virasoro constraint
in \p{Virasoro}) in the formalism made us introduce the zweibein
$E_\alpha{}^+$ in the otherwise flat derivatives \p{3.8}.
\q{III.2}

\subsection{The twistor superstring action}

The twistor superstring action consists of three terms,
\be
S= S_1 + S_2 + S_3.
\ee
The first of them is a replica of the superparticle action \p{2.21}
\be\q{3.15}
S_1 = \int {d^2x d^8\theta}\;
\left[iP_{\alpha\un a } E_\alpha{}^{\un a } +
P_{\alpha\beta\un a } E_{\{\alpha}\gamma^{\un a} E_{\beta\}}\right].
\ee
Its invariance under the worldsheet diffeomorphisms of subsection
\ref{III.2} is once again due to the homogeneous transformation laws for
the spinor derivatives $D_\alpha$.  The meaning of the two equations of
motion introduced by Lagrange multipliers is the same as in the
superparticle case, and so is the component content of the action
\p{3.15} (cf. \p{compact}):
\be\q{3.16}
S_1 = \int d^2x\; \left[p_{\un a} \left({\cal E}_-{}^{\un a} - {1\over 8}
{\lambda}_\alpha \gamma^{\un a} {\lambda}_\alpha \right) +
p_{\alpha\beta\un a}
\lambda_{\{\alpha}\gamma^{\un
a} \lambda_{\beta\}} \right].
\ee
Note that the vector ${\cal E}_-{}^{\un a}={E}_-{}^{\un a}\vert_0$
contains the covariant derivative \p{3.14} (in the Wess-Zumino gauge
\p{3.13}).
Eliminating the twistor variables from \p{3.16} by varying with respect to
the Lagrange multipliers $p_{\un a}$ and $p_{\alpha\beta\un a}$, we obtain
\be\q{3.17}
S_1 = \int d^2x\; g_{++}\; {\cal E}_-{}^{\un a}{\cal E}_{-\un a},
\ee
where $g_{++}(x)$ is a new Lagrange multiplier, needed to impose the
lightlike condition on the vector ${\cal E}_-{}^{\un a}$,
\be\q{3.18}
{\cal E}_-{}^{\un a}{\cal E}_{-\un a} = 0.
\ee
Actually, this is one of the two Virasoro constraints for the string, so
the Lagrange multiplier $g_{++}$ is the second component of the worldsheet
metric (the first one is $g_{--}$ from \p{3.14}).

The second term in the superstring action is
\be\q{enf}
S_2 = \int d^2x d^8\theta P_{\alpha\beta}D_{\{\alpha} E_{\beta\}}{}^+ .
\ee
It enforces the constraint \p{3.9} on the only non-trivial zweibein
$E_\alpha{}^+ $ on shell. This allows one to treat it as an unconstrained
superfield off shell. The solution to this constraint (in the Wess-Zumino
gauge \p{3.13}) was found in
\p{3.14}, and one easily sees that the only essential component $g_{--}$ of
$E_\alpha{}^+$ does not appear in $S_2$. Therefore $S_2$ is a purely
auxiliary term in the action, and does not contribute to the component
superstring action.

The last and most interesting term in the twistor superstring action
involves the two-form $B_{\un{MN}}(\un z)$ of background supergravity. It
is constructed in close analogy with the superparticle-Maxwell coupling
term \p{m7} of subsection
\ref{II.6}. The Lagrange multiplier in that term yielded the equation of
motion \p{m8}, which meant that the pull-back of the Maxwell one-form
$A_{\un M}$ to the worldsheet was a pure gauge. The integrability
condition \p{m6}, which made that possible, followed from the constraints
\p{m2} on the field strength. Finally, on shell the Lagrange multiplier
itself reduced to a constant, the electric charge. In the case of the
supergravity two-form we shall follow the same strategy.

The two-form $B_{{\un{M}}\un{N}} $ is defined up to abelian gauge
transformations,
\be\q{ab}
\delta B_{{\un{M}}\un{N}}  = \pl_{\un{M}}
b_{\un{N}} (\un z) - (-)^{{\un{M}}\un{N}} \pl_{\un{N}} b_{\un{M}} (\un z)
{}.
\ee
Consequently, the pull-back
\be\q{pull}
B_{ M N} =(-)^{M(N+{\un N})}
\pl_{ N}z^{\un{N}}  \pl_{ M }z^{\un{M}} B_{\un{M}{\un{N}}}
\ee
undergoes the transformations
\be\q{abp}
\delta B_{ M  N} = \pl_{ M } b_{ N} - (-)^{ M  N}
\pl_{ N} b_{ M }, \ \ \ b_{ M } = \pl_{ M }z^{\un{M}}
b_{\un{M}} (\un z) .
\ee
Following the analogy with eq. \p{m8}, we could try to make $B_{ M N}$
pure gauge on shell,
\be\q{3.19}
B_{ M N} = \pl_{[M} Q_{N\}},
\ee
where $[\}$ means graded antisymmetrisation. However, this would put too
strong a restriction on the background. The integrability condition for
\p{3.19} is
\be\q{3f}
0 = \pl_{[ M }B_{ N K\}} = H_{ M N K} = (-)^{(M+N)(K+{\un K}) +M(N+{\un
N})}\pl_{ K}z^{\un K}
\pl_{ N}z^{\un{N}}  \pl_{ M }z^{\un{M}}
H_{\un{M}\un N\un K}(\un z)
\ee
$$ \rightarrow \ \ H_{ABC} =(-)^{(A+B)(C+K)+A(B+N)} E_C{}^KE_B{}^NE_A{}^M
H_{MNK} = 0.  $$ One can show that it is not compatible with the
supergravity constraints
\p{3.3}, \p{3.4} on the three-form. Indeed, using the equation of motion
\p{2.12} coming from the term \p{3.15} of the superstring action
\footnote{Using an equation of motion produced by a Lagrange multiplier in
some other term of the action is equivalent to finding an appropriate
redefinition of that Lagrange multiplier.} and applying the constraint
\p{3.3}, one finds that
\be\q{3.20}
H_{\alpha\beta\gamma} = 0.
\ee
Further, the constraint \p{3.4} together with \p{2.10} and
\p{2.20} imply
\be\q{3.21}
H_{\alpha\beta-} = -2iE_{\alpha}{}^{\un\alpha} E_\beta{}^{\un\beta}
E_-{}^{\un a} (\gamma_{\un a} )_{\un{\alpha\beta}} \;e^{\Phi} =
-2i\delta_{\alpha\beta} E_{-{\un a} } E_-{}^{\un a} \;e^\Phi = 0.
\ee
Similarly,
\be\q{3.22}
H_{\alpha\beta+} = {1\over 8} \delta_{\alpha\beta} H_{\gamma\gamma+}
=-2i\delta_{\alpha\beta} E_-{}^{\un a} E_{+{\un a} }\;e^\Phi \neq 0.
\ee
The last component of $H_{ A B C}$ can be found from the Bianchi identity
\be\q{bia}
(D_{(\alpha}+ \pl_+E_{(\alpha}{}^+)H_{\beta\gamma)+} -
\pl_+H_{\alpha\beta\gamma}  =-2i \delta_{(\alpha\beta} H_{\gamma)-+}.
\ee
With the help of \p{3.20} and \p{3.22} this gives
\be\q{bi}
H_{\alpha-+} = {i\over 16} (D_\alpha + \pl_+E_\alpha{}^+)H_{\beta\beta+}
\neq 0.
\ee

We see that the components \p{3.22} and \p{bi} of the three-form pull-back
do not vanish, so the integrability condition \p{3f} is too strong.  This
suggests that the correct on-shell condition on the two-form pull-back
should be somewhat weaker than \p{3.19}. The right choice is the ``almost
flat" pull-back \footnote{Note that if the pull-back $B_{MN}$ is
restricted to the ``--" and ``$\mu$" directions, eq. \p{em} becomes
completely flat. This is another manifestation of lightlike integrability,
which in this case applies to the left-handed sector of the worldsheet
only.}
\be\q{em}
B_{ M N} + E_{[ M }{}^+E_{ N\}}{}^- e^\Phi E_-{}^{\un a} E_{+{\un a} } =
\pl_{[ M } Q_{ N\}},
\ee
where $E_M{}^\pm$ are elements of the zweibein matrix on the worldsheet.
The integrability condition for \p{em} is
\be\q{3.23}
H_{KMN}+ \pl_{[K} \left(E_{ M }{}^+E_{ N\}}{}^- e^\Phi E_-{}^{\un a}
E_{+{\un a} }\right) = 0.
\ee
In a tangent space basis it reads
\be\q{eq}
H_{ A B C} = {i\over 16}(-1)^{(A+B)(C+K)+A(B+N)} E_{[ C}{}^{ K } E_{
B}{}^{N} D_{ A\}}\left(E_{ N }{}^+E_{ K}{}^-H_{\delta\delta+} \right),
\ee
where the term with $e^\Phi$ has been replaced by the three-form component
from \p{3.22}.  To check that \p{eq} holds one should use the expressions
\p{3.20}, \p{3.21}, \p{3.22}, the
identity \p{bi} and the constraint \p{3.9}.

As in the case of the superparticle-Maxwell coupling, the on-shell
``almost flat" condition
\p{em} will be obtained from a Lagrange multiplier term in the superstring
action:
\be\q{3.24}
S_3 = \int d^2x d^8\theta \; \left[B_{ M N} + E_{[ M }{}^+E_{ N\}}{}^-
e^\Phi E_-{}^{\un a} E_{+{\un a} } - \pl_{[ M } Q_{ N\}} \right] P^{MN} .
\ee
It is invariant under the diffeomorphisms of subsection \ref{III.2}.
Indeed, the terms with $B$ and $Q$ transform as curved worldsheet
supertensors. The same applies to the term with $\Phi$. The easiest way to
see this is as follows. Using \p{2.12} and \p{2.20}, one can complete the
trace in the term $E_M{}^+ E_{+\un a} = E_M{}^A E_{A\un a} = E_{M\un a}$,
so this term is covariant.  \footnote{At this point it becomes clear why
we did not need to covariantise the derivative $D_+$ in
\p{3.8}. The only occurrence of a worldsheet tangent index $``+"$ in the
action is in the term $E_M{}^+ E_{+\un a}$, and the presence of any
zweibeins $E_+{}^N$ in it would be irrelevant, as we have just seen.} The
other term, $E_N{}^- E_-{}^{\un a}$, is covariant as well, since
$E_-{}^{\un a}$ transforms into itself, $\delta E_-{}^{\un a} = -
(D_-\Lambda^- )E_-{}^{\un a}$ (see
\p{tr-}
and use \p{2.12}), and the zweibein $E_N{}^-$ compensates for this.

To find the component content of the action term \p{3.24} we have to
repeat the argument of subsection \ref{II.6}. We first look at the
equations of motion following from the variation with respect to the
Lagrange multiplier $Q_M$:
\be\q{q}
\pl_N P^{NM} = 0.
\ee
It is easy to see that this equation has the general solution
\be\q{sol}
P^{MN} = \pl_K\Sigma^{KMN} + \theta^8 \delta_+{}^{[M}\delta_-{}^{N\}}\; T,
\ee
where $\Sigma^{KMN}(z)$ is a totally (graded) antisymmetric superfield and
$T$ is a constant,
\be\q{tens}
\pl_+ T = \pl_- T =  0.
\ee
The origin of the cohomology term with $T$ in \p{sol} can be traced back
to the $``+"$ and $``-"$ projections of \p{q}:
\be
\pl_- P^{-+} + \pl_\mu P^{\mu +} = 0, \ \ \ \pl_+ P^{+-} + \pl_\mu P^{\mu
-} = 0 .
\ee
In both of them the second terms contain the odd derivative $\pl_\mu$,
therefore the $\theta^8$ term is missing in their expansions. This leads
to the constraints \p{tens} on the highest-order term in $P^{+-}$.

Further, the arbitrary superfield $\Sigma$ in \p{sol} actually corresponds
to a gauge symmetry of the action. This follows from the integrability
condition \p{3.23}.  Therefore one can gauge away almost everything in
$P^{MN}$ but its highest-order {\it constant} component.  \footnote{It is
not hard to see that this is a gauge of the Wess-Zumino type, i.e. no
parameters with space-time derivatives are used.} After this has been
done, the component form of $S_3$ can be obtained very easily:
\bea
S_3 &=& \int d^2x d^8\theta\; \theta^8 T(B_{+-} + e^\Phi E_+{}^{\un a}
E_{-\un a})\vert_0 \nn\\ &=& T\int d^2x \; (\pl_-z^{\un N}\pl_+z^{\un M}
B_{\un{MN}} + e^\Phi {\cal E}^{\un a}_+ {\cal E}_{-\un a}), \q{f}
\eea
where $\un z$ denotes the lowest-order component of the superfield $\un
z(z)$.  Clearly, this term has retained its original symmetries, notably
the local $N=(8,0)$ worldsheet supersymmetry. Of course, in \p{f} we have
assumed that $T\neq 0$, otherwise the action would be trivial.

Finally, we put together the component expressions of the terms $S_1$
\p{3.17} and $S_3$ \p{f}, redefine the Lagrange multiplier $g_{++}\rightarrow
Te^\Phi g_{++}$ and insert the expression for the covariant derivative
$D_-$ in the Wess-Zumino gauge \p{3.14}. The result is
\bea\q{3.100}
S &=& T \int d^2x \left\{\pl_-z^{\un N}\pl_+z^{\un M}B_{\un{MN}}
\right. \\ &+& e^\Phi \left. \left[{\cal
E}_+{}^{\un a}{\cal E}_{-{\un a}}(1+2g_{++}g_{--}) + {\cal E}_-{}^{\un
a}{\cal E}_{-{\un a}}g_{++} + {\cal E}_+{}^{\un a}{\cal E}_{+{\un
a}}g_{--}(1+g_{++}g_{--})\right]\right\} , \nn
\eea
where ${\cal E}_\pm = \pl_\pm z^{\un M} E_{\un M}{}^{\un a}\vert_0$ now
involve only partial derivatives $\pl_\pm$.  Up to a conventional
redefinition of the two-dimensional metric, the action \p{3.100} is that
of the Green-Schwarz superstring \p{3.2}.

At this point we can show the main difference between our approach to the
superstring and that of Tonin \cite{TON}. He does not use the non-trivial
mechanism of generating the string tension as an integration constant on
shell, but introduces it by hand. This prevents him from writing down a
worldsheet superspace Wess-Zumino term. Instead, he proposes an
essentially component term similar to eq. \p{f}, thus loosing manifest
worldsheet supersymmetry.  Further, to prove supersymmetry he is forced to
put gauge non-covariant restrictions directly on the two-form. As we have
shown, using the Wess-Zumino term
\p{3.24} of the Chern-Simons type
allowed us to maintain all the symmetries manifest, and at the same time
to clearly exhibit the geometry of the superstring action.

\section{Conclusions}

In the present paper we developed a new geometric formulation of the
heterotic superstring in ten dimensions. Its main feature is that all the
inherent symmetries are linearly realised and manifest. In particular,
kappa-symmetry of the usual Green-Schwarz formulation is replaced by
worldsheet supersymmetry. We also established the equivalency between the
two formulations at the classical level.  An issue of crucial importance
is whether the new action is can be covariantly quantised, thus hopefully
overcoming the well-known difficulties in the Lorentz-covariant
quantisation of the Green-Schwarz action.

A further open question is how to generalise the twistor approach to
superstrings of the non-heterotic type, which have full $N=(8,8)$
worldsheet supersymmetry, as well as to other super p-branes. This would
involve dealing with other dimensions of the target space, notably $D=11$.
It is not clear at present if the twistor-like approach can be extended
beyond the special dimensions $D=3,4,6,10$.

To complete the heterotic superstring one needs a matter sector of chiral
fermions (or bosons). At the moment we do not know how to formulate it,
keeping {\it both} $N=(8,0)$ worldsheet supersymmetry and target space
Lorentz symmetry manifest and having a compact matter Yang-Mills group.
Perhaps trying to couple the superstring to an external super-Maxwell
field may provide the key to this problem.  Another, probably related task
is to study in detail the meaning and implications of the principle of
lightlike integrability in the new context (throughout the text we have
made several remarks about that). It is not impossible that the new
construction may also inspire a solution of the long-standing problem of
off-shell super-Yang-Mills and supergravity in ten dimensions.

\vskip15mm
{\bf Acknowledgements} During the work on this paper E.S. and F.D.
profited from visits to ENS Lyon and the University of Bonn, respectively,
in the framework of the European programme PROCOPE. P.H. thanks P.Sorba
for hospitality at ENS Lyon.  A.G. and E.S. would like to thank M.Tonin
for sending us his very interesting preprint \cite{TON}. E.S. is grateful
to R.Flume for carefully reading the manuscript.

\end{document}